\documentclass[prc,twocolumn,nofootinbib,superscriptaddress,showpacs]{revtex4}
\usepackage{graphicx,epstopdf,amsmath,amssymb,bm,multirow}
\usepackage{subfig}
\usepackage{hhline}

\newcommand{\be}[1]{\begin{equation}\label{#1}}
\newcommand{\ee}{\end{equation}}

\newcommand{\ket}[1]{|#1\rangle}

\begin{document}

\title{Alternative similarity renormalization group generators in nuclear structure calculations}

\author{Nuiok M. Dicaire}
\email[E-mail:~]{ndica015@uottawa.ca}
\affiliation{Department of Physics, University of Ottawa, Ottawa, ON, K1N 6N5, Canada}
\affiliation{TRIUMF, 4004 Wesbrook Mall, Vancouver, BC, V6T 2A3, Canada}
\author{Conor Omand}
\email[E-mail:~]{comand92@hotmail.com}
\affiliation{Department of Physics and Astronomy, University of British Columbia, Vancouver, BC, V6T 1Z4, Canada}
\affiliation{TRIUMF, 4004 Wesbrook Mall, Vancouver, BC, V6T 2A3, Canada}
\author{Petr Navr\'atil}
\email[E-mail:~]{navratil@triumf.ca}
\affiliation{TRIUMF, 4004 Wesbrook Mall, Vancouver, BC, V6T 2A3, Canada}

\begin{abstract}
The similarity renormalization group (SRG) has been successfully applied to soften interactions for {\it ab initio} nuclear calculations. In almost all practical applications in nuclear physics, an SRG generator with the kinetic energy operator is used. With this choice, a fast convergence of many-body calculations can be achieved, but at the same time substantial three-body interactions are induced even if one starts from a purely two-nucleon ($NN$) Hamiltonian. Three-nucleon ($3N$) interactions can be handled by modern many-body methods. However, it has been observed that when including initial chiral $3N$ forces in the Hamiltonian, the SRG transformations induce a non-negligible four-nucleon interaction that cannot be currently included in the calculations for technical reasons. Consequently, it is essential to investigate alternative SRG generators that might suppress the induction of many-body forces while at the same time might preserve the good convergence. In this work we test two alternative generators with operators of block structure in the harmonic oscillator basis.
In the no-core shell model calculations for $^3$H, $^4$He and $^6$Li with chiral $NN$ force, we demonstrate that their performances
appear quite promising.
\end{abstract}


\maketitle


\section{Introduction}
\label{Sec:Intro}

One of the major goals of nuclear physics is to understand the structure and dynamics of nuclei, i.e. quantum many-body systems exhibiting bound states, unbound resonances, and scattering states, all of which can be strongly coupled. {\it Ab initio} (i.e., from first principles) approaches attempt to achieve such a goal starting from accurate basic interactions among nucleons. The modern theory of nuclear forces based on the chiral effective field theory ($\chi$EFT) \cite{Epelbaum_2009,Machleidt:2011zz} offers a link to the underlying theory of quantum chromodynamics at low energies. Nucleon-nucleon ($NN$) and three-nucleon ($3N$) interactions derived with the help of the chiral EFT have been recently used with a significant success as input to various many-body techniques. Methods such as the no-core shell model (NCSM)~\cite{Navratil:2000ww,Barrett:2013nh}, coupled cluster (CC) theory~\cite{Hagen:2012fb,Hagen:2013nca,Binder:2012mk,Binder:2013xaa}, self-consistent Green's functions (SCGF)~\cite{Cipollone:2013zma,Soma:2013xha}, in-medium SRG~\cite{Hergert:2012nb,Hergert:2013uja} and lattice EFT~\cite{Epelbaum:2012qn,Epelbaum:2013paa,Lahde:2013uqa} calculate binding energies, excitation energies, separation energies, radii, transition rates, and other observables for light as well as medium mass nuclei. They provide tests of these forces and at the same time provide predictions that can be confronted with experiments. Some of these methods can also be extended to describe resonances, scattering states and even nuclear reactions, e.g., no-core shell model with continuum (NCSMC)~\cite{Baroni:2012su,Baroni:2013fe} or CC with the Gamow basis~\cite{Hagen:2012rq}.

Even though the chiral interactions are in general softer than the traditional $NN$ potentials constructed from the meson-exchange theory, they still pose convergence problems for the many-body methods. Only CC calculations for closed shell nuclei were performed to convergence using bare chiral $NN$ potentials~\cite{Hagen:2012fb,Ekstrom:2013kea}. None of the above many-body methods, however, is at present capable of reaching sufficiently large model spaces when the chiral $NN$ interaction is supplemented by the chiral $3N$ interaction. Consequently, techniques such as the similarity renormalization group (SRG) have been applied to soften the chiral interactions~\cite{Glazek93,Wegner94,Furnstahl:2012fn,Furnstahl:2013oba}. 
The SRG uses continuous series of unitary transformations of the free-space Hamiltonian $H$ ($H{\equiv}H_{s=0}$),
\begin{equation}
H_{s} = U_{s} H U_{s}^{\dagger} \;,
\label{eq:Hs}
\end{equation}
to decouple high-momentum and low-momentum physics.
The label $s$ is a flow parameter running from zero toward $\infty$, which keeps track of the sequence of Hamiltonians.
These transformations are implemented as a flow equation in $s$
\begin{equation}
 \frac{dH_{s}}{ds} = [\eta_s,H_s] = [[G_{s},H_{s}],H_{s}]  \;,
 \label{eq:flow}
\end{equation}
whose form guarantees that the $H_s$'s are unitarily equivalent~\cite{Bogner:2006pc,Wegner94,Kehrein:2006ti}.
Here, $\eta_s=\frac{dU}{ds}U^\dagger$ is an anti-hermitian SRG generator chosen in a form $\eta_s=[G_s,H_s]$ with a hermitian flow operator $G_s$.
The high- and low-momentum decoupling results in general in a faster convergence of many-body calculations. At the same time, the SRG transformation induces many-body forces, i.e., even if the initial $H_{s=0}$ Hamiltonian includes only two-body interactions, the evolved $H_{s>0}$ Hamiltonians will contain many-body interactions, in principle up to $A$-body for an $A$-nucleon system.

In {\it ab initio} nuclear calculations, the SRG generator has been typically chosen by setting $G_s{=}T_{\rm rel}$, where the $T_{\rm rel}$ is the relative kinetic energy operator~\cite{Furnstahl:2012fn,Furnstahl:2013oba}. With this choice, the convergence is fast and the evolution of many-body forces can be consistently performed~\cite{Bogner:2006pc}. By varying the flow parameter $s$ and using it as a gauge of the unitarity of SRG transformations, it has been found that starting from a Hamiltonian with a chiral $NN$ interaction, there are significant induced three-body forces, but the induced four- and higher-body interactions appear negligible~\cite{Jurgenson:2009qs,Jurgenson:2010wy,Roth:2011ar}. As the $3N$ interactions can be handled by modern many-body methods, the SRG transformations of Hamiltonians with chiral $NN$ interactions facilitate the solution of the quantum many-body problem for light and medium mass nuclei. However, it has been observed that when the initial chiral 3N forces are present in the Hamiltonian, the SRG transformations induce non-negligible four-nucleon interactions for systems with $A{\gtrsim}10$ ~\cite{Roth:2011ar,Roth:2011vt} that cannot be currently included in the calculations for technical reasons. The problem can be circumvented to some extent by a reduction of the cutoff of the initial chiral $3N$ interaction~\cite{Roth:2011vt}, but such a solution is far from satisfactory as, e.g., it limits the parameter space of the chiral forces and the range of applicability of these forces. One might expect that in heavier nuclei higher momenta might become more important than in light systems and therefore higher cutoffs of nuclear forces could be appropriate. In general, the issue of consistent choices of $NN$ and $3N$ interactions, regularization schemes, and cutoffs is open, see, e.g., the discussion in the recent Ref.~\cite{Hagen:2013yba}. The N$^3$LO $3N$ contributions will soon be tested in many-body calculations, one might anticipate the problem of the SRG four- and higher-body induced interactions to re-emerge. 

The strength of the induced many-body interactions and the rate of convergence depend on the choice of the SRG generator. Consequently, it is essential to investigate alternative SRG generators to the standard choice of $G_s{=}T_{\rm rel}$ that might suppress the induction of many-body forces while at the same time might preserve the good convergence. In this work we test two alternative generators with operators $G_s$ of block structure in the harmonic oscillator (HO) basis. We evolve the chiral $NN$ interaction of Refs.~\cite{Machleidt:2011zz,Entem:2003ft} using these novel generators and apply them in NCSM nuclear structure calculations for $^3$H, $^4$He and $^6$Li. In this initial study, we limit ourselves to the SRG $NN$-only interactions and demonstrate good convergence properties as well as a reduction of the induced three- and higher many-body forces compared to the standard kinetic-term generator.

In Sect.~\ref{Sec:Formalism}, we introduce the tested alternative generators and provide a brief description of the NCSM approach. Our results are summarized in Sect.~\ref{Sec:Results}. Conclusions and outlook are given in Sect.~\ref{Sec:Conclusions}.

\section{Formalism}
\label{Sec:Formalism}

\subsection{Background}
\label{Subsec:Background}

The starting Hamiltonian of {\it ab initio} approaches can be written as
\begin{eqnarray}\label{ham}
H= T_{\rm rel} + {\cal V} &=&
\frac{1}{A}\sum_{i<j}\frac{(\vec{p}_i-\vec{p}_j)^2}{2m} + \sum_{i<j}^A V_{{\rm NN}, ij} \nonumber \\
&&+ \sum_{i<j<k}^A V_{{\rm NNN}, ijk} \; ,
\end{eqnarray}
where $m$ is the nucleon mass, $V_{{\rm NN}, ij}$ is the $NN$ interaction, and $V_{{\rm NNN}, ijk}$ is the $3N$ interaction. The
$T_{\rm rel}{=}\frac{1}{A}\sum_{i<j}\frac{(\vec{p}_i-\vec{p}_j)^2}{2m}$ is the intrinsic kinetic energy of the $A$-nucleon system. In the present work, we employ the chiral N$^3$LO $NN$ interaction of Refs.~\cite{Machleidt:2011zz,Entem:2003ft} that fits two-nucleon scattering data accurately up to $\approx 300$ MeV. We omit the chiral $3N$ interaction in this initial study.

The SRG transformation can be performed systematically starting from the two-nucleon system, then proceeding to the three-nucleon system, etc. Embedding the SRG evolved $NN$ interaction in the three-nucleon space allows isolation of the pure three-nucleon part. Similarly, the SRG evolved two-nucleon and the three-nucleon interaction can be embedded in the four-nucleon space and the SRG evolved four-nucleon part can be isolated, etc., although in practice the procedure has been so far performed only up to the three-nucleon level (note, however, that first attempts to calculate and apply SRG induced four-nucleon interactions have been already done~\cite{Roth}). This procedure is particularly well established for the $G_s{=}T_{\rm rel}$ choice of the generator~\cite{Bogner:2006pc}.

The SRG evolution of the $A{=}2$ system is typically performed in the momentum space~\cite{Bogner:2006pc,Furnstahl:2012fn,Furnstahl:2013oba,Roth:2010bm}. On the other hand, the SRG evolution of the $A{=}3$ system has been first accomplished using the HO basis~\cite{Jurgenson:2009qs,Jurgenson:2010wy} although later a momentum space implementation~\cite{Hebeler:2012pr} and also hyperspherical momentum implementation~\cite{Wendt:2013bla} have been achieved.

While the original choice for $G_s$ advocated by Wegner and collaborators~\cite{Wegner94,Kehrein:2006ti} and applied extensively in condensed matter is the diagonal component of the Hamiltonian $G_s{=}H^d_ s$, in most practical applications of the SRG in nuclear physics, the $G_s{=}T_{\rm rel}$ choice in the generator was used. However, there were several exploratory studies focusing on alternative generator choices. As to the use of the diagonal generator in nuclear physics see Ref.~\cite{Wendt:2011qj}. Then in Ref.~\cite{Anderson:2008mu}, a block diagonal operator
\begin{equation}
 \begin{aligned}
G_{s} = \begin{pmatrix} P_{\Lambda}H_s P_{\Lambda}&0\\ 0&Q_{\Lambda}H_s Q_{\Lambda} \end{pmatrix}  
 \end{aligned}
\label{eqn:momblock}
\end{equation}
in the two-nucleon momentum space was introduced and tested in nucleon-nucleon phase shift calculations. In a partial wave momentum representation, the projection operators $P$ and $Q=1{-}P$ are step functions defined by a sharp cutoff $\Lambda$ on relative momenta. With this choice, the SRG evolved $NN$ potential is loosely related to the low momentum interaction $V_{{\rm low} {\it k}}$ constructed by a renormalization group method by preserving the two-nucleon T matrix~\cite{Bogner:2002yw,Bogner:2003wn}.

A variation of the standard kinetic operator choice was explored in Ref.~\cite{Li:2011sr} where the $G_s$ was chosen as functions of $T_{\rm rel}$. In particular, $G_s=-\frac{\sigma^2}{1+T_{\rm rel}/\sigma^2}$ and $G_s=-\sigma^2 e^{-T_{\rm rel}/\sigma^2}$ were considered with a parameter $\sigma$ controlling the separation of a low-momentum region. It was demonstrated that with these generators the computational time is reduced and, at the same time, the low- and high-momentum separation can be tailored to some extent by particular choices of $\sigma$. However, these generators were only tested in two-nucleon calculations and in few-body calculations in a one-dimensional model, i.e., no realistic nuclear calculations with $A{>}2$ were performed with these or any other alternative generators.

\subsection{Block SRG generators}
\label{Subsec:BlockGen}

\begin{figure}%
    \centering
    \subfloat{{\includegraphics[width=7cm]{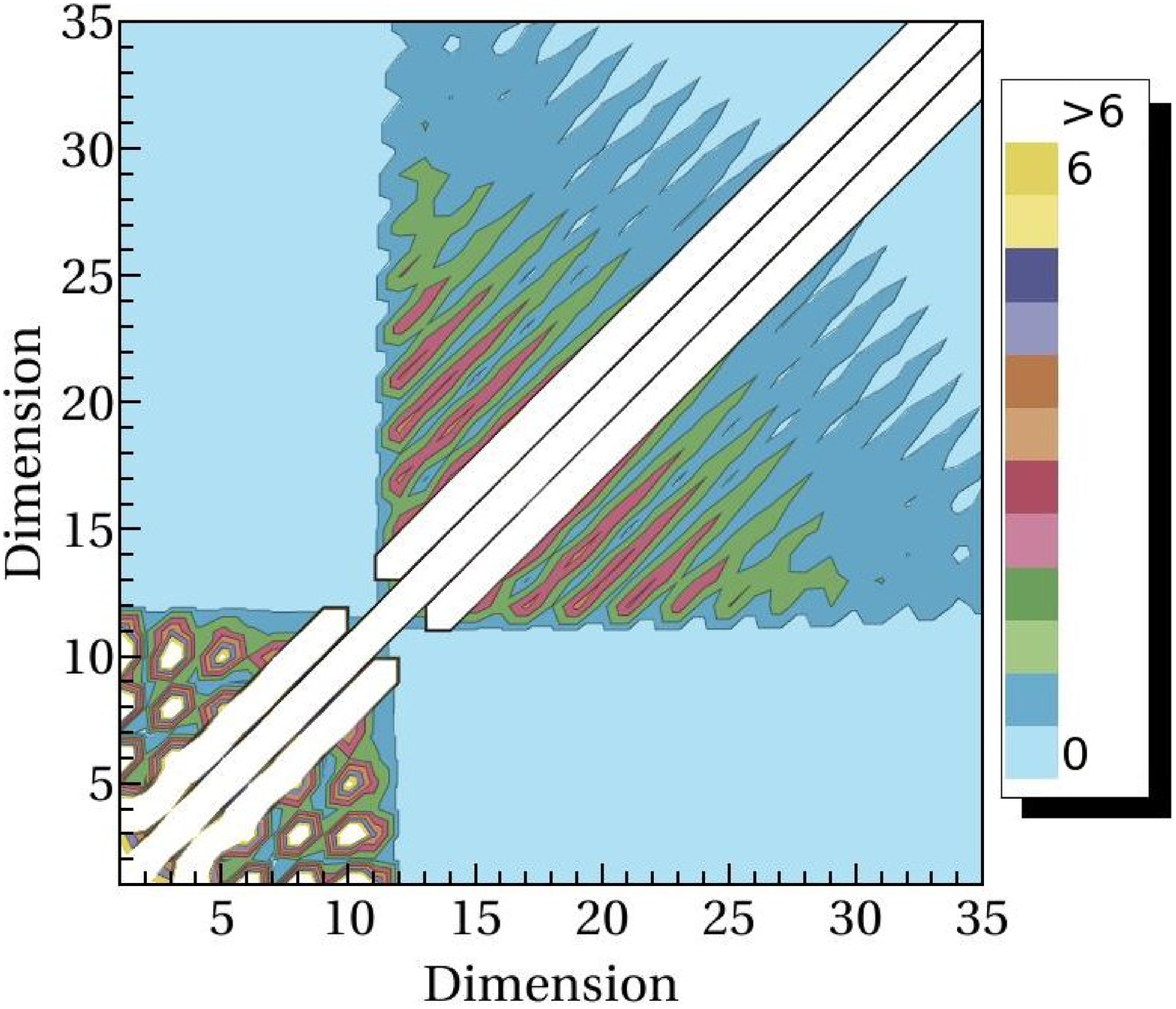} }}%
    \qquad
    \subfloat{{\includegraphics[width=7cm]{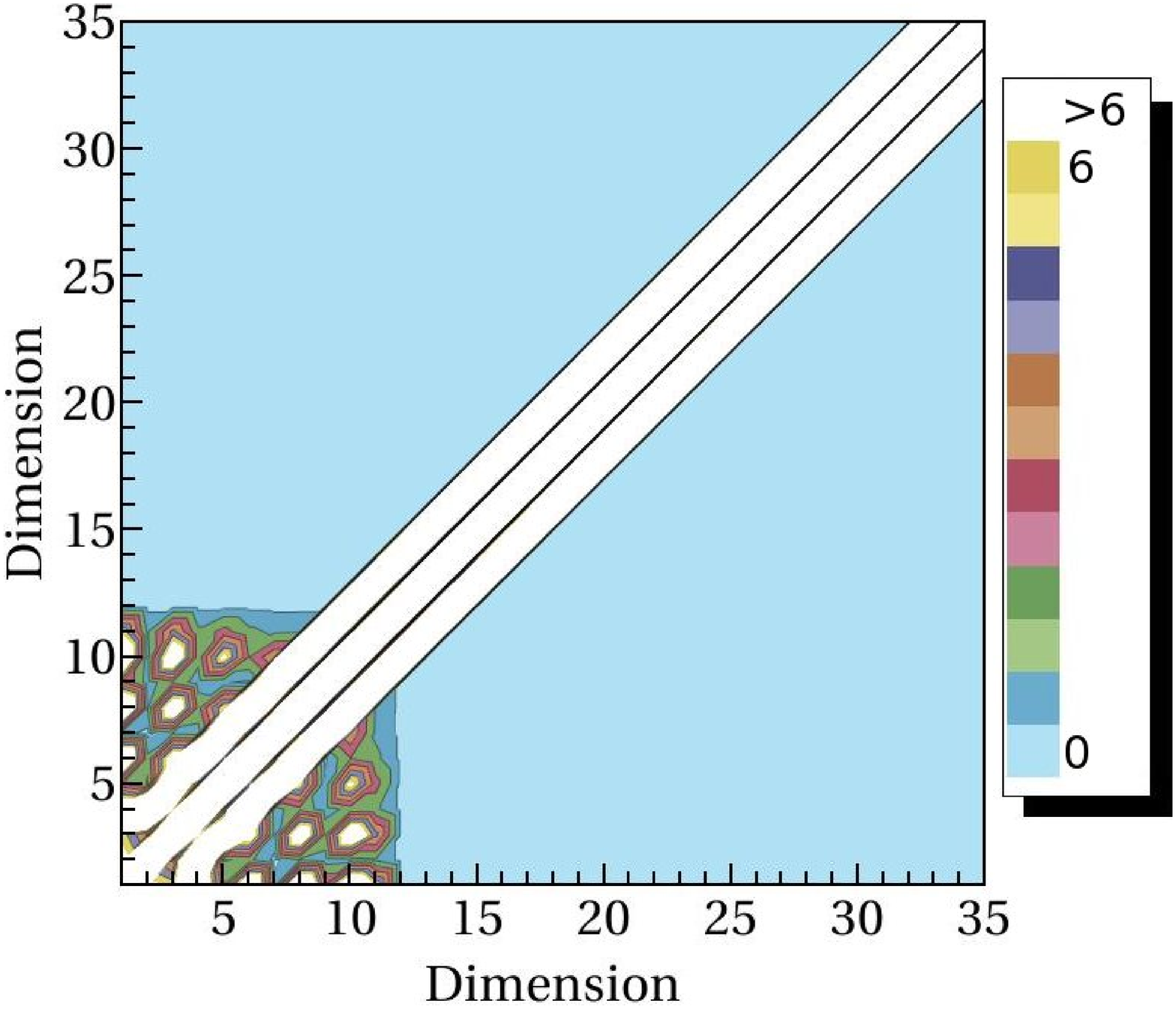} }}%
    \caption{(color online) Absolute values of matrix elements of the flow operator $G_{s=0}^A$ (top) and $G_{s=0}^B$ (bottom) in the relative-coordinate HO basis of the $^3S_1{-}^3D_1$ $NN$ channel. The $N_{\rm gen}{=}10$ was used with the corresponding size of the model space 11 and the HO frequency of $\hbar\Omega=16$ MeV. The three white bands along the diagonal correspond to large matrix elements of the kinetic operator. The initial ($s{=}0$) chiral N$^3$LO $NN$ interaction of Refs.~\cite{Machleidt:2011zz,Entem:2003ft} was used. Units are in MeV.}%
    \label{fig1a}%
\end{figure}
We introduce and test in few nucleon calculations two alternative generators with the flow operator $G_s$ of block structure in the HO basis. As most of the {\it ab initio} many-body methods employ the HO basis in one way or another, it is sensible to consider generators with cuts in the HO basis. While the following discussion is mostly general, we note that in this paper we perform only the SRG evolution of $NN$ interactions in the two-nucleon space and then apply the evolved $NN$ interactions in many-nucleon calculations.

First, let us define an SRG generator with the $G_s$ as the block-diagonal part of H$_s$ in analogy with Eq.~(\ref{eqn:momblock}) but in the HO, rather than the momentum basis, i.e.,
\begin{equation}
 \begin{aligned}
G_{s}^A = \begin{pmatrix} P_{\rm gen}H_s P_{\rm gen}&0\\ 0&Q_{\rm gen}H_s Q_{\rm gen} \end{pmatrix} \\ P_{\rm gen}:N\leq N_{\rm gen}\ \ \ \ \ \ Q_{\rm gen}:N> N_{\rm gen}
 \end{aligned}
\label{eqn:Ablock}
\end{equation}
with $N$ the number of HO excitations of all nucleons above the minimum configuration in $A$-nucleon basis states. See the panel (a) of Fig.~\ref{fig1a} for a schematic representation. As $T_{\rm rel}$ is band-diagonal in the HO basis and drives the Hamiltonian to be band-diagonal, $G_{s}^A$ should drive the Hamiltonian to be block-diagonal in the spirit of the Okubo-Lee-Suzuki approach~\cite{Okubo:1954zz,Suzuki:1980yp,Suzuki:1982aa_JPV,Suzuki:1983aa_JPV} used, e.g., in earlier NCSM studies~\cite{Navratil:2000ww}. We let $N_{\rm gen}$, which determines the size of $P_{\rm gen}$, be independent of the definition of the $A$-nucleon basis space defined, e.g., in the NCSM calculations by $N\leq N_{\rm max}$.

Second, let's consider a generator with the flow operator $G_s$ given by (see Fig.~\ref{fig1a}, panel (b))
\begin{equation}\label{eqn:Bblock}
G_s^B = T_{\rm rel} +  P_{\rm gen}V_s P_{\rm gen}\; , \;\; P_{\rm gen}:N\leq N_{\rm gen} \;.
\end{equation}
With this choice, a Hamiltonian with a potential with zero matrix elements for basis states with $N{>}N_{\rm gen}$ will not be transformed. A realistic potential with this property is, e.g., the inverse-scattering JISP $NN$ interaction~\cite{Shirokov:2003kk}. More generally, if a (starting) potential has only weak matrix elements in the basis states with $N{>}N_{\rm gen}$, it will be only mildly affected by the SRG transformation generated by the flow operator $G_s^B$. As long as we can reach a basis with, e.g., $N_{\rm max}\geq N_{\rm gen}$, we can solve the many-nucleon problem of such a Hamiltonian within, e.g., the NCSMC method~\cite{Baroni:2013fe}. It is then counterproductive to SRG transform such a Hamiltonian using, e.g., the standard $G_s{=}T_{\rm rel}$ and generate many-body terms in the process. SRG transformations generated with $G_s^B$ (and also by $G_s^A$) will in general modify the initial Hamiltonian less than the $G_s{=}T_{\rm rel}$ transformations, i.e., one may hope to induce weaker many-body forces. Further, the $G_s^B$, unlike the $G_s^A$~(\ref{eqn:Ablock}), will not strive to eliminate the strong $T_{\rm rel}$ matrix elements that couple $N_{\rm gen}$ with $N_{\rm gen}{+}2$ basis states.

We note that both $G_s^A$ and $G_s^B$ now depend on the HO frequency $\Omega$. Consequently, the evolved Hamiltonian will no longer be variational with respect to $\Omega$.  However, these generators may suppress the induction of three- and  higher-body terms, making it easier to preserve the unitarity of the transformation while being computationally easier.

\subsection{NCSM}
\label{Subsec:NCSM}

To test the alternative SRG generators, we perform NCSM calculations for light nuclei. The {\it ab initio} NCSM is a technique appropriate for
the description of bound states or for approximations of narrow resonances. With the Hamiltonian given by Eq.~(\ref{ham}), nuclei are considered as systems of $A$ non-relativistic point-like nucleons interacting through realistic inter-nucleon interactions, i.e., those that describe accurately two-nucleon and, possibly, three-nucleon systems. All nucleons are active degrees of freedom. Translational invariance as well as angular momentum and parity of the system under consideration are conserved. The many-body wave function is cast into an expansion over a complete set of antisymmetric $A$-nucleon HO basis states containing up to $N_{\rm max}$ HO excitations above the lowest possible configuration:
\begin{equation}\label{NCSM_wav}
 \ket{\Psi^{J^\pi T}_A} = \sum_{N=0}^{N_{\rm max}}\sum_i c_{Ni}\ket{ANiJ^\pi T}\; .
\end{equation}
Here, $N$ denotes the total number of HO excitations of all nucleons above the minimum configuration,  $J^\pi T$ are the total angular momentum, parity and isospin, and $i$ additional quantum numbers. The sum over $N$ is restricted by parity to either an even or odd sequence. The basis is further characterized by the frequency $\Omega$ of the HO well and may depend on either Jacobi relative or single-particle coordinates. In the former case, the wave function does not contain the center of mass (c.m.) motion, but antisymmetrization is complicated~\cite{Navratil:1999pw}. In the latter case, antisymmetrization is trivially achieved using Slater determinants, but the c.m.\ degrees of freedom are included in the basis. The HO basis  within the $N_{\rm max}$ truncation is the only possible one that allows an exact factorization of the c.m.\ motion for the eigenstates, even when working with single-particle coordinates and Slater determinants. Calculations performed with the two alternative coordinate choices are completely equivalent~\cite{Barrett:2013nh}.

\subsection{Parameters}
\label{Subsec:Param}

To assess the performance of the alternative generators, we perform the SRG transformations in the $A{=}2$ system using the HO basis (with $N_{\rm max}\approx 300$ sufficient for convergence). Then we apply the SRG evolved $NN$ interaction in $A{>}2$ NCSM calculations. The initial chiral $3N$ interaction as well as the SRG induced $3N$ interaction is neglected in this first study. It will be a subject of a future investigation. We compare results obtained using the block generators with flow operators $G_s^A$ (\ref{eqn:Ablock}) and $G_s^B$ (\ref{eqn:Bblock}) to the standard generator with $G_s{=}T_{\rm rel}$.

In our study, we vary the following four parameters:
\begin{itemize}
\item $N_{\rm gen}$: The total number of HO excitations that defines the projector P$_{\rm gen}$ and sets the dimension of the blocks of $G_s^A$ and $G_s^B$.
\item $s$: (or $\lambda\equiv 1/s^{1/4}$) Sets the degree of SRG evolution (larger $s$, smaller $\lambda$, is more evolved).
\item $N_{\rm max}$: The total number of HO excitations used in the NCSM calculation.
\item $\hbar\Omega$: HO basis parameter that controls the shape of the HO potential well and the eigenenergies.
\end{itemize}
and examine the effects on the Hamiltonian, convergence, and calculated binding energies in $^3$H, $^4$He, and $^6$Li.

\section{Results}
\label{Sec:Results}

\subsection{SRG evolved $NN$ potentials}
\label{Subsec:Potentials}

\begin{figure*}[tbh]
\begin{minipage}{40pc}
  \centering
  \includegraphics[width=40pc]{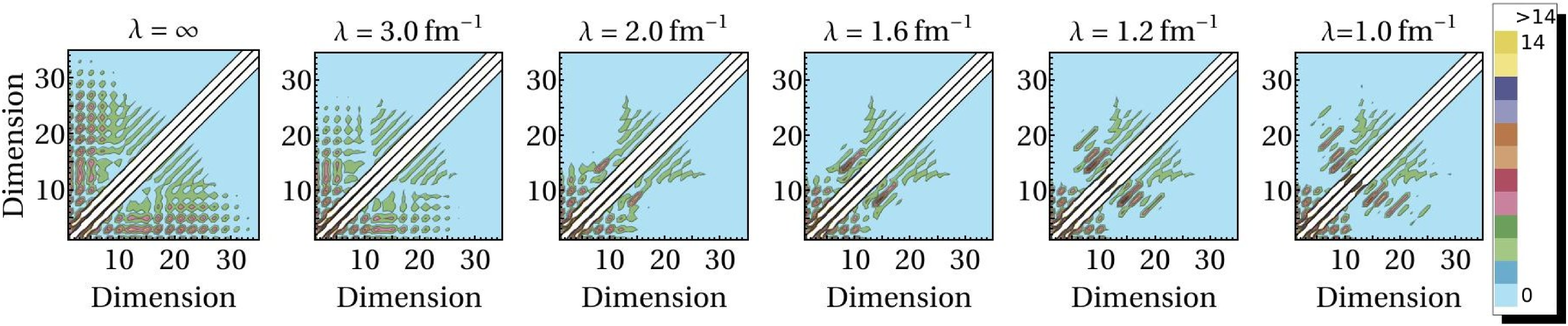}
\end{minipage}\hspace{3pc}%
\begin{minipage}{40pc}
    \includegraphics[width=40pc]{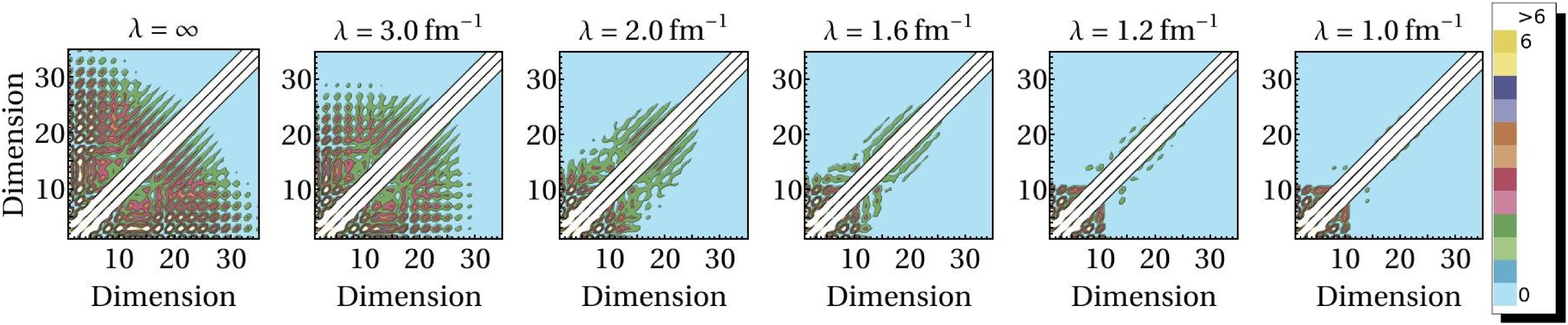}
\end{minipage}
\caption{(color online) Absolute values of the SRG evolved two-nucleon Hamiltonian matrix elements (in MeV) in the relative-coordinate HO basis of the $^3S_1{-}^3D_1$ $NN$ channel for (top row) Generator A  and (bottom row) Generator B for various degrees of evolution. The $N_{\rm gen}{=}10$ was used with the corresponding size of the model space 11 and the HO frequency of $\hbar\Omega=16$ MeV. The initial ($s{=}0$) chiral N$^3$LO $NN$ interaction of Refs.~\cite{Machleidt:2011zz,Entem:2003ft} was used.}
\label{fig4}
\end{figure*}
\begin{figure}[t]
\begin{center}
\includegraphics[clip=,width=0.45\textwidth]{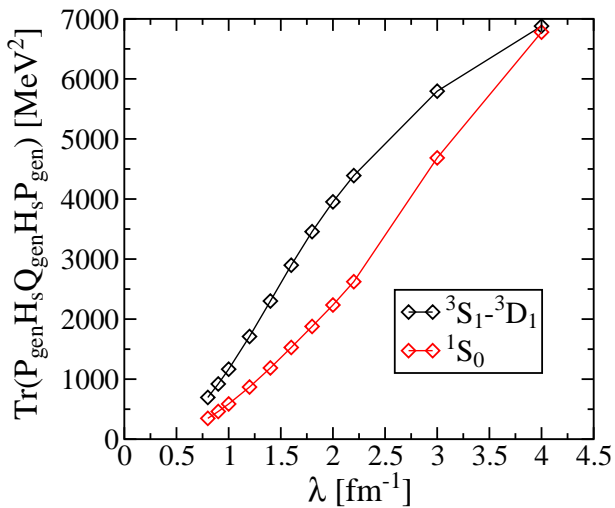}
\end{center}
\caption{(color online) The decoupling measure ${\rm Tr}(P_{\rm gen}H_sQ_{\rm gen}H_sP_{\rm gen})$ dependence on the SRG evoution parameter $\lambda$ for the $^3S_1-^3D_1$ and the $^1S_0$ $NN$ channels. The Generator A and the parameters as described in Fig.~\protect\ref{fig4} (top row) were used.}
\label{sum_mat}
\end{figure}
\begin{figure*}[tbh]
\begin{minipage}{40pc}
  \centering
  \includegraphics[width=40pc]{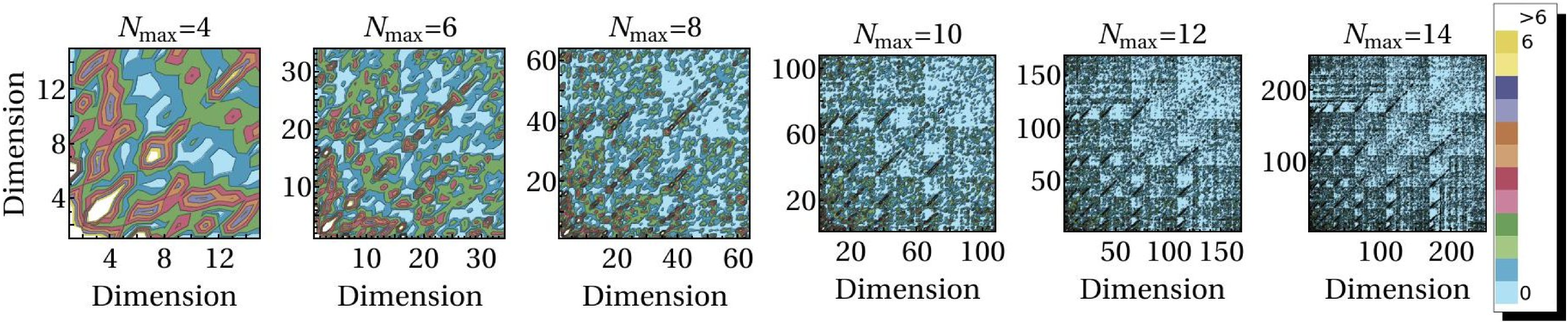}
\end{minipage}\hspace{3pc}%
\begin{minipage}{40pc}
    \includegraphics[width=40pc]{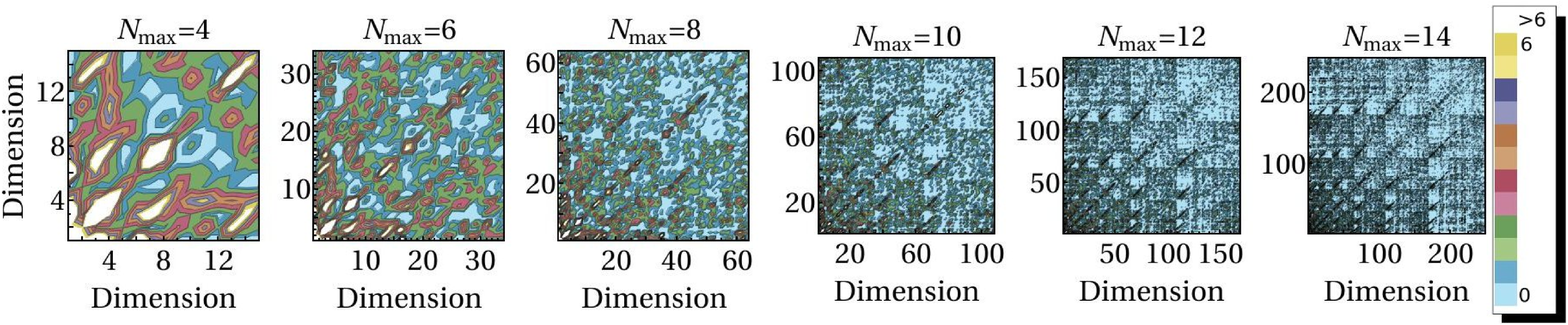}
\end{minipage}
\caption{(color online) Absolute values of the $NN$ potential matrix elements (in MeV) in the antisymmetrized Jacobi-coordinate three-nucleon HO basis for the $J^\pi T{=}\frac{1}{2}^+ \frac{1}{2}$ $^3$H channel for various sizes of the three-nucleon model space. Bare initial $NN$ interaction (top row) and the SRG evolved $NN$ interaction with the generator B (bottom row) with $N_{\rm gen}{=}10$, HO frequency of $\hbar\Omega{=}24$ MeV and the SRG flow parameter $\lambda=2.0$ fm$^{-1}$ were used.}
\label{fig6}
\end{figure*}
\begin{figure}[t]
\begin{center}
\includegraphics[clip=,width=0.45\textwidth]{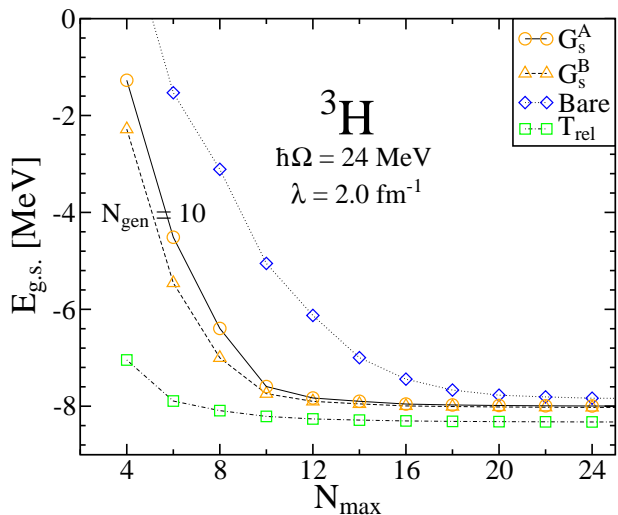}
\end{center}
\caption{(color online) The $^3$H ground-state energy dependence on the size of the basis using the SRG evolved $NN$ potentials presented in Fig.~\protect\ref{fig6}. For a comparison, calculations with the (bare) initial chiral N$^3$LO $NN$ interaction and with the SRG evolved $NN$ using the $G_s{=}T_{\rm rel}$ flow operator are also shown.}
\label{Hfig0}
\end{figure}

\begin{figure*}[tbh]%
    \centering
    \subfloat{{\includegraphics[width=8cm]{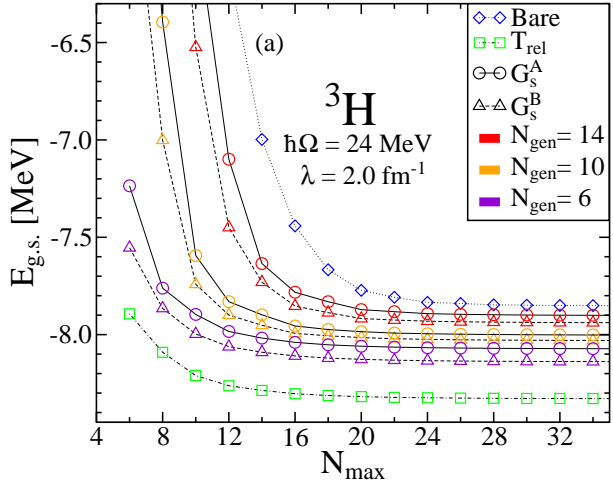} }}%
    \qquad
    \subfloat{{\includegraphics[width=8cm]{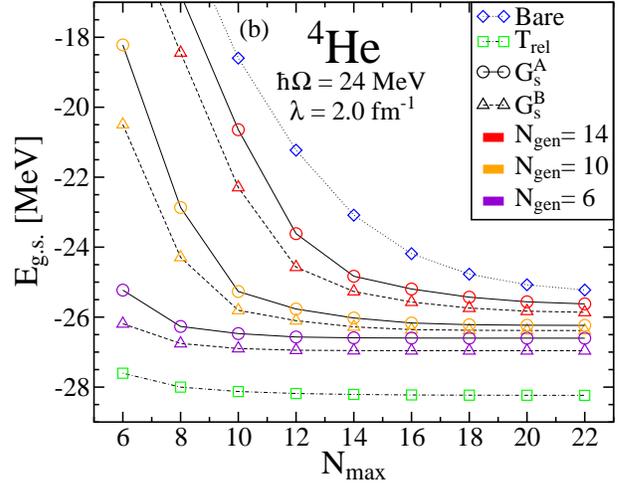} }}%
    \caption{(color online) Figure (a) shows the ground-state energy for $^3$H as a function of $N_{\rm max}$, for generators A and B at various $N_{\rm gen}$, as well as for the bare initial interaction and for when the kinetic operator was used as the flow operator. The frequency is set to 24 MeV and $\lambda$ is set to $2.0$ fm$^{-1}$. (b) Same as above, but for $^4$He.} %
    \label{Hfig1}%
\end{figure*}

In Fig.~\ref{fig4} we show the evolved $NN$ Hamiltonians in the Jacobi coordinate two-nucleon HO basis of the $^3S_1{-}^3D_1$ $NN$ channel. The dimension of the model space $P_{\rm gen}$ is equal to 11, when using generators A and B with $N_{\rm gen}{=}10$ in the SRG transformation. The Hamiltonians are shown at different degree of the evolution (i.e. different values of the flow parameter $s$ or $ \lambda$). The evolved Hamiltonians depend on the dimension of the model space $P_{\rm gen}$, which is determined by the parameter $N_{\rm gen}$, and to some extent also on the HO frequency.

The evolved $NN$ Hamiltonians obtained with both generators are similar at high $\lambda$ (low $s$). In both cases we see that the Hamiltonians are driven to a block-diagonal form as we expected. With generator A, when evolving below $\lambda{\sim}1.6$ fm$^{-1}$ it appears that some off-diagonal matrix elements are induced and the matrices have a less pronounced block structure. This is attributed to the design of this generator, i.e., to the fact that the large coupling elements of the kinetic operator at the boundary of the $P_{\rm gen}$ and $Q_{\rm gen}$ spaces are removed from the flow operator, see the top panel of Fig.~\ref{fig1a}. This in turn eliminates the strong coupling elements of the kinetic term at the boundary as visible in particular in the figure for $\lambda{\sim}1.0$~fm$^{-1}$ and in the process induces some weaker off-diagonal matrix elements. We do not eliminate these boundary coupling kinetic term matrix elements when using the generator B. The resulting evolved $NN$ Hamiltonian in that case has a clearly visible block structure and a narrow diagonal part in the $Q_{\rm gen}$ space already at intermediate degrees of evolution.  

It should be noted, however, that also in the case of the Generator A (top row of Fig.~\ref{fig4}), the $P_{\rm gen}H_sQ_{\rm gen}$ and $Q_{\rm gen}H_sP_{\rm gen}$ blocks do get systematically eliminated with increasing $s$ (decreasing $\lambda$) as anticipated based on general arguments presented in Ref.~\cite{Anderson:2008mu} (see Eq. (4) there). We demonstrate this in Fig.~\ref{sum_mat} that presents the decoupling measure  ${\rm Tr}(P_{\rm gen}H_sQ_{\rm gen}H_sP_{\rm gen})$ monotonically decreasing with decreasing $\lambda$ for two $NN$ channels. The deuteron $^3S_1-^3D_1$ channel results correspond to those plotted in the top row of Fig~\ref{fig4}.

\subsection{$^3$H and $^4$He}
\label{Subsec:3H+4He}

Figure~\ref{fig6} presents the bare initial and the SRG evolved $NN$ potentials in the antisymmetrized Jacobi-coordinate three-nucleon HO basis for the $J^\pi T{=}\frac{1}{2}^+ \frac{1}{2}$ $^3$H channel for various sizes of the three-nucleon model space characterized by $N_{\rm max}$. In Fig.~\ref{Hfig0}, we show the corresponding $^3$H ground-state energy as a function of the parameter $N_{\rm max}$. A subtle block structure is visible in all plots showing the potential, although clearly not of the type seen in the two-nucleon basis (Fig.~\ref{fig4}). Although we do not see significant distinguishable differences due to the SRG evolution, the plots showing the energy as a function of $N_{\rm max}$ clearly give different energies depending on the generator that is used, proving that differences are present in the $NN$ potentials. Fig.~\ref{Hfig0} demonstrates that convergence is somewhat slower for the generators A and B compared to the standard  $G_s{=}T_{\rm rel}$, with the converged ground-state energies being much closer to the bare potential result obtained with the bare initial $NN$ potential. Clearly, the convergence is much faster with any of the SRG evolved $NN$ potential compared to the bare one.

\begin{figure}[t]
\begin{center}
\includegraphics[clip=,width=0.45\textwidth]{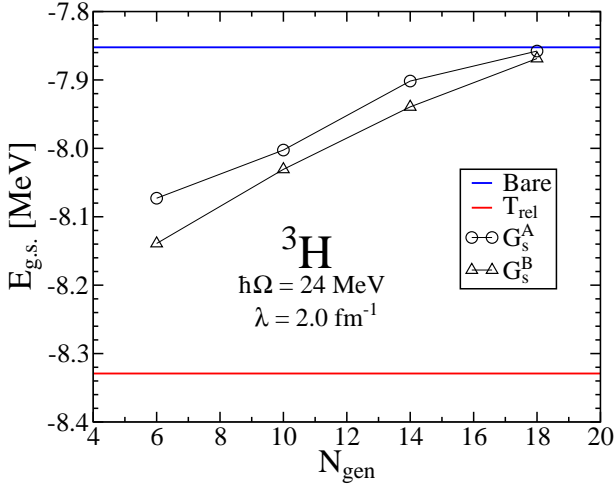}
\end{center}
\caption{(color online) Ground-state energy of $^3$H as a function of $N_{\rm gen}$ with $\hbar\Omega{=}24$ MeV and $\lambda=2.0$ fm$^{-1}$. Results obtained with the bare initial $NN$ potential and the SRG evolved with $G_s{=}T_{\rm rel}$ are shown as full lines.}
\label{Hfig7}
\end{figure}
\begin{figure}[tbh]
\begin{center}
\includegraphics[clip=,width=0.45\textwidth]{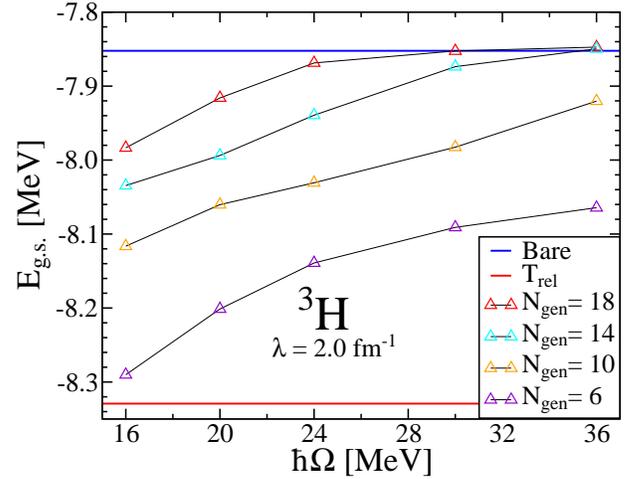}
\end{center}
\caption{(color online) Ground-state energy of $^3$H as a function of HO frequency for different values of $N_{\rm gen}$ using $G_s^B$ with $\lambda{=}2.0$ fm$^{-1}$.}
\label{Hfig6}
\end{figure}

Figures~\ref{Hfig1} (a) and (b) also present ground-state energies obtained after evolving the $NN$ potentials with SRG transformations. In particular, Figure~\ref{Hfig1} (a) examines the energy dependence as a function of the basis dimension, $N_{\rm max}$, for $^3$H while Figure~\ref{Hfig1} (b) presents the same results for $^4$He. We show the curves obtained for the two block structured generators. In order to characterize these new generators we also include as a mean of reference the ground-state energies obtained by the bare initial interaction and the ones obtained by using the kinetic term $G_s{=}T_{\rm rel}$ as the flow operator for the SRG evolution. The frequency is set to 24 MeV and the evolution is performed up to $\lambda=2.0$~fm$^{-1}$. The values obtained with the kinetic term, $G_s{=}T_{\rm rel}$, as the flow operator show the largest binding energy for both $^3$H and $^4$He, and are farther away from the values obtained for the bare interaction than any of the values obtained with generator A or B. Nevertheless, this generator gives an extremely quick convergence, which makes it, to this day, the most widely used generator when SRG techniques are employed in nuclear physics. We seek to examine the convergence properties of the two generators proposed in this study, as well as their induced higher-body components to access their usefulness in {\it ab initio} nuclear structure calculations. We first note that in both cases, as the dimension of the $P_{\rm gen}$ model increases (i.e. as $N_{\rm gen}$ increases) there is less induced higher-body terms, but the energy only converges at a larger $N_{\rm max}$. Thus using very large values of $N_{\rm gen}$ yields the same values as the bare interaction, while very small values of $N_{\rm gen}$ give results similar to using $G_s{=}T_{\rm rel}$. This is clearly demonstrated in Fig.~\ref{Hfig7}, where converged $^3$H ground-state energies are shown for various $N_{\rm gen}$ values. Overall, the energies obtained for generator A and for generator B are somewhat similar to each other. Generator A consistently yields states that are slightly less bound than generator B which implies that less three-body forces are induced. This energy difference is more pronounced before convergence is achieved and also increases with $N_{\rm gen}$ before convergence. The converged values obtained for $^3$H do not differ by much when using one generator or the other. Although the overall trends observed for $^4$He are the same, since this nucleus is more tightly bound we observe firstly that the overall energies are greater and secondly that there is a larger difference between the converged energy given by the kinetic term and the ones from the bare interaction.

As already noted, the generators A and B that we introduced depend also on the HO frequency. In Fig.~\ref{Hfig6}, we show the converged $^3$H ground-state energies obtained with the generator B for a wide range of HO frequencies and $N_{\rm gen}$ values. In general, the bigger the $N_{\rm gen}\hbar\Omega$ product, the fewer $NN$ repulsive short range correlations are transformed away by the SRG transformation, which in turn results in less binding of nuclei. The same trend is also observed for generator A in $^3$H and $^4$He calculations. The $G_s{=}T_{\rm rel}$ and the bare interaction converged results are both frequency independent. Consequently the $G_s{=}T_{\rm rel}$ SRG transformation is variational in HO frequency, while with generators A and B it is not.

\begin{figure*}[tbh]
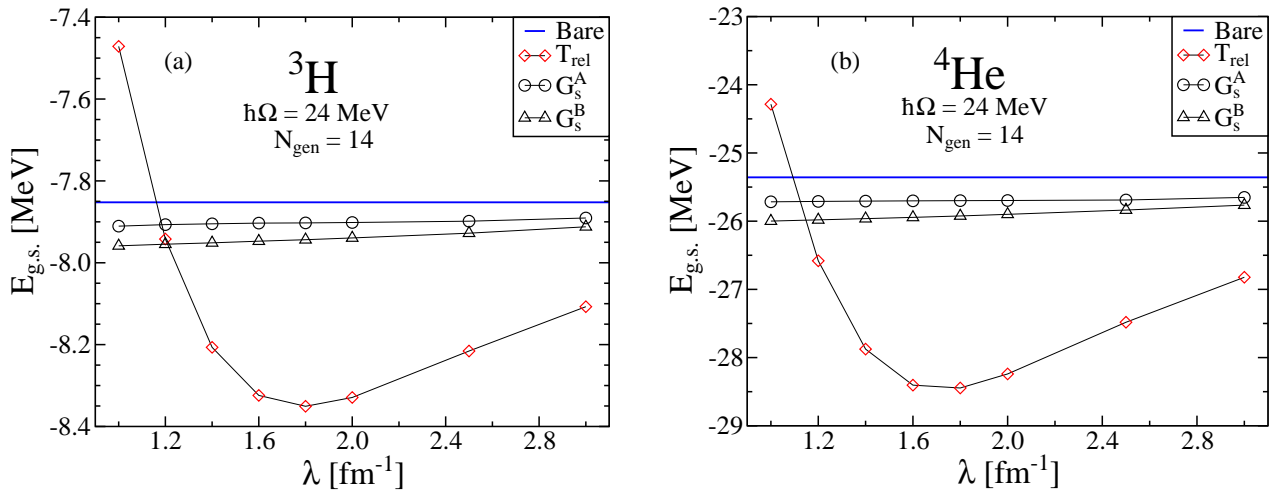
%
    \centering
    \subfloat{{\includegraphics[width=8cm]{Hfig5.eps} }}%
    \qquad
    \subfloat{{\includegraphics[width=8cm]{Hfig4.eps} }}%
    \caption{(color online) Ground-state energy of $^3$H (a) and $^4$He (b) as a function of the flow parameter for the three different generators as well as the ground-state energy obtained using the bare initial interaction. For generators A and B, $N_{\rm gen}{=}14$ and the HO frequency of $\hbar\Omega{=}24$ MeV was used. All energy values are converged or extrapolated.} %
    \label{Hfig4}%
\end{figure*}
\begin{table*}[tbh]
\begin{center}
\begin{ruledtabular}
\begin{tabular}{ccccc||ccccc}
$^3$H &  $\lambda$ [fm$^{-1}$]  &  $N_{gen}$   &  $\hbar\Omega$ [MeV]  &  $E_{\rm g.s.}$ [MeV]  & $^4$He & $\lambda$ [fm$^{-1}$]  &  $N_{gen}$   &  $\hbar\Omega$ [MeV]  &  $E_{\rm g.s.}$ [MeV]  \\
\hline \hline 
Bare & - & - & 24 & -7.85 & Bare & - & - & 24 & -25.39 \\
$T_{\rm rel}$ & 2.0 & - & 24 & -8.33 & $T_{\rm rel}$ & 2.0 & - & 24 & -28.24 \\
\hline
$G_s^A$ & 2.0 & 6 & 24 & -8.07 & $G_s^A$ & 2.0 & 6 & 24 & -26.60\\ 
 & 2.0 & 10 & 24 & -8.00 & & 2.0 & 10 & 24 & -26.24\\
 & 2.0 & 14 & 24 & -7.90 & & 2.0 & 14 & 24 & -25.62\\
 & 2.0 & 18 & 24 & -7.86 & & 2.0 & 18 & 24 & -25.29\\
 & 1.6 & 14 & 24 & -7.90 & & 1.6 & 14 & 24 & -25.63\\
 & 1.2 & 14 & 24 & -7.91 & & 1.2 & 14 & 24 & -25.64\\
 & 2.0 & 14 & 36 & -7.81 & & 2.0 & 14 & 36 & -25.37\\
 & 2.0 & 14 & 16 & -8.01 & & 2.0 & 14 & 16 & -26.23\\
\hline
$G_s^B$ & 2.0 & 6 & 24 & -8.14 & $G_s^B$ & 2.0 & 6 & 24 & -26.96 \\
 & 2.0 & 10 & 24 & -8.03 & & 2.0 & 10 & 24 & -26.38 \\
 & 2.0 & 14 & 24 & -7.94 & & 2.0 & 14 & 24 & -25.86 \\
 & 2.0 & 18 & 24 & -7.87 & & 2.0 & 18 & 24 & -25.39 \\
 & 1.6 & 14 & 24 & -7.95 & & 1.6 & 14 & 24 & -25.91 \\
 & 1.2 & 14 & 24 & -7.95 & & 1.2 & 14 & 24 & -26.96 \\
 & 2.0 & 14 & 36 & -7.81 & & 2.0 & 14 & 36 & -25.37 \\
 & 2.0 & 14 & 16 & -8.04 & & 2.0 & 14 & 16 & -26.38 \\
\end{tabular}
\end{ruledtabular}
\caption{Ground-state energies of $^3$H and $^4$He for the different flow operators and various choices of parameters. 
}
\label{tbl2}
\end{center}
\end{table*}

Previous studies~\cite{Jurgenson:2009qs,Jurgenson:2010wy,Roth:2011ar,Roth:2011vt} analyzed the lambda dependence in the converged energy when using the kinetic term as the flow operator. They showed a significant $\lambda$ dependence in calculations with $NN$-only interactions due to the fact that the low and the high momentum matrix elements are affected by the SRG transformation at different stages of the evolution because some of the information is transferred into $3N$ terms that were not taken into account. However this dependence was shown to be mostly removed when including the $3N$ interactions in the calculations, in particular when no initial chiral $3N$ interactions were included. In Figure \ref{Hfig4}, we examine the $^3$H and $^4$He ground-state energy variation as we evolve the Hamiltonians. The energies correspond to the converged values obtained for a given $N_{\rm gen}$ and a given HO frequency. Both generator A and generator B show basically no $\lambda$ dependence for $^3$H and very little such dependence in the case of $^4$He. This is contrary to the $G_s{=}T_{\rm rel}$ case where, at first, the ground-state energy decreases as the repulsive short range part of the $NN$ is transformed away and later it increases when the attractive part of the $NN$ begins to be removed. The block generators that we introduced do not significantly affect the medium range attractive part of the $NN$ potential and remove a smaller part of the short range repulsion. Consequently, the lambda dependence alone, with the selected choice of the $N_{\rm gen}$ and $\hbar\Omega$, is much weaker and the binding energy larger. 

It should be noted, however, that the meaning of $\lambda$ (or $s$) is different for the present alternative generators compared to the $G_s{=}T_{\rm rel}$ case. For the latter, $\lambda$ is a measure of the width of the band in momentum representation, and therefore sets the scale for the momentum transfers which occur when particles interact via the SRG-evolved $NN$ potential. For the block-diagonal decoupling $G^A_s$, on the other hand, the evolution should in principle be carried out all the way to infinity, i.e., $\lambda$ driven all the way to zero. The scale of the interaction is set by the choice of the $P_{\rm gen}$ and $Q_{\rm gen}$ operators, specifically by $N_{\rm gen}\hbar\Omega$. For the $G^B_s$ flow operator, the choice of $P_{\rm gen}$, and therefore $N_{\rm gen}\hbar\Omega$ also sets the physically relevant scale, although $\lambda$ remains a measure for the width of the band of matrix elements in the $Q_{\rm gen}$ space. Consequently, when judging the energy variations due to the SRG evolution using generators A and B, one should look simultaneously at the $\lambda$, $N_{\rm gen}$ and $\hbar\Omega$ dependencies (Figs.~\ref{Hfig7}, \ref{Hfig6}, and \ref{Hfig4}). We then conclude that the variations are similar as in the $G_s{=}T_{\rm rel}$ case. 

\begin{figure}[tbh]
\begin{center}
\includegraphics[clip=,width=0.45\textwidth]{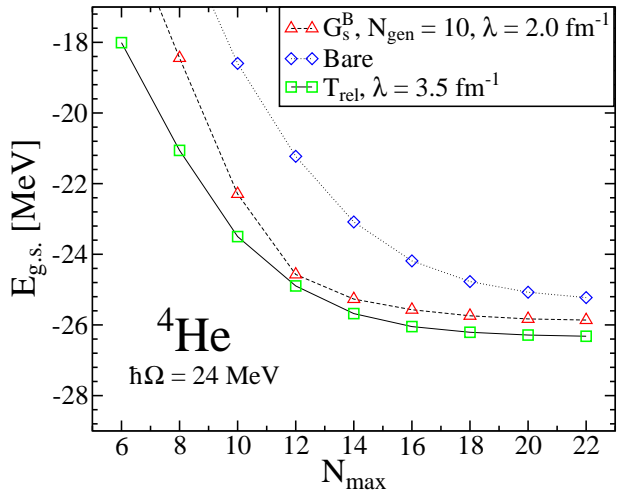}
\end{center}
\caption{(color online) The bare $NN$ and the $G^B_s$ $N_{\rm gen}{=}14$ results from Fig.~\protect\ref{Hfig1} (b) compared to the $G_s{=}T_{\rm rel}$ and $\lambda=3.5$~fm$^{-1}$ calculation.}
\label{Hfig8}
\end{figure}
To get still a deeper insight into the comparison between the $G_s{=}T_{\rm rel}$ and the present alternative generators, we note that the ultraviolate (UV) cutoff for the latter can be estimated as $\Lambda_{\rm UV}{\sim}\sqrt{m_N/\hbar^2} \sqrt{(N_{\rm gen} {+} 3/2) \hbar\Omega}$. For the parameters used in Fig.~\ref{Hfig1}, this gives $\Lambda_{\rm UV}{\sim} 3$~fm$^{-1}$ for the $N_{\rm gen}{=}14$. In Fig.~\ref{Hfig8}, we re-plot the bare and $G^B_s$ $N_{\rm gen}{=}14$ results for $^4$He from Fig.~\ref{Hfig1} (b) and include for a comparison a calculation with the $G_s{=}T_{\rm rel}$ and $\lambda{=}3.5$~fm$^{-1}$ (the $G_s{=}T_{\rm rel}$ result with the $\lambda{=}3$~fm$^{-1}$ is shwon in Fig.~\ref{Hfig4} (b); it is below the $G^B_s$ one). From Fig.~\ref{Hfig8}, we can see that the convergence of the $G^B_s$ calculation is comparable or even faster once $N_{\rm max}$ is greater than $N_{\rm gen}$ and induced many-body forces are somewhat weaker than those obtained with the $G_s{=}T_{\rm rel}$ with a comparable UV cutoff.

We present a sample of our calculated $^3$H and $^4$He ground-state energies in Table~\ref{tbl2}.

\subsection{$^6$Li}
\label{Subsec:6Li}

We also performed similar calculations for the more complex nucleus $^6$Li. The $NN$ interactions were SRG evolved in the HO Jacobi basis and then transformed into a HO Slater determinant (SD) basis used typically in all NCSM calculations with $A{>}4$. The SD basis simplifies the antisymmetrization of the wave function that becomes prohibitively difficult in Jacobi coordinates when using a larger number of nucleons, $A$.
\begin{figure}[tbh]
\begin{center}
\includegraphics[clip=,width=0.45\textwidth]{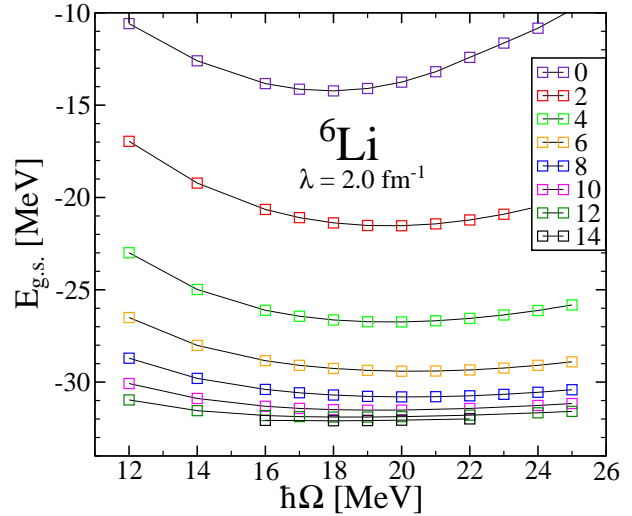}
\end{center}
\caption{(color online) Ground-state energy as a function of HO frequency using the $G_s{=}T_{\rm rel}$ flow operator for different values of $N_{\rm max}$ with $\lambda{=}2.0$ fm$^{-1}$.}
\label{Lifig1}
\end{figure}
We first consider the frequency dependence of the $^6$Li ground-state energy when the $G_s{=}T_{\rm rel}$ is used as the generator of the SRG transformation. These results are presented in Fig.~\ref{Lifig1}. Since the use of a larger basis exponentially increases the computation time, we were generally limited to values of $N_{\rm max}$ up to $12$ or $14$. Note that even for the highest basis size, in this case $N_{\rm max}{=}14$, the calculation has not yet fully converged but an accurate extrapolated value can be obtained, as described below. In this figure, we also observe a minimum in energy for each value of $N_{\rm max}$ although this minimum varies and tends towards smaller frequencies as $N_{\rm max}$ increases. Because the kinetic energy is frequency independent, the variational principle stipulates that the frequency with a minimum in energy should give the closest approximation of the true value. Moreover, since higher $N_{\rm max}$ values are closer to a converged value, we can adopt the frequency at the highest $N_{\rm max}$ as giving the best approximation, which in this case is $18$ MeV.

\begin{figure}[tbh]
\begin{center}
\includegraphics[clip=,width=0.45\textwidth]{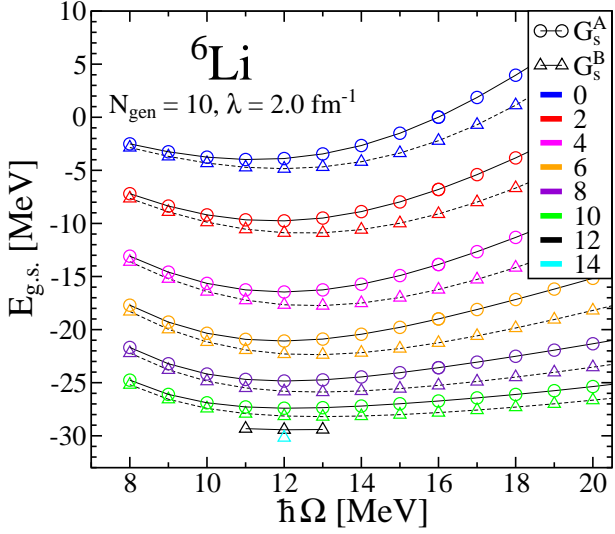}
\end{center}
\caption{(color online) Ground-state energy as a function of HO frequency using generator A and B for different values of $N_{\rm max}$ with $\lambda{=}2.0$ fm$^{-1}$ and $N_{\rm gen}{=}10$.}
\label{Lifig2}
\end{figure}
\begin{figure}[tbh]
\begin{center}
\includegraphics[clip=,width=0.45\textwidth]{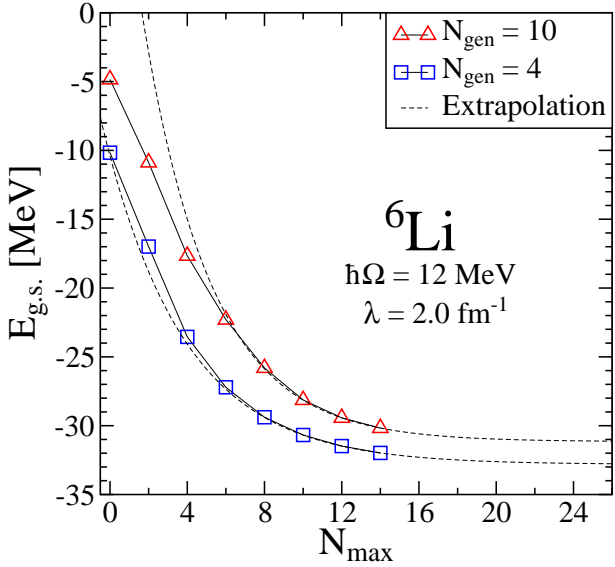}
\end{center}
\caption{(color online) Ground-state energy as a function of $N_{\rm max}$ for generator B using $\lambda{=}2.0$ fm$^{-1}$ and $N_{\rm gen}{=}10$. Results are shown for the minimum frequency of 12 MeV associated with this specific set of parameters. The extrapolated curve is also shown.}
\label{Lifig3}
\end{figure}
We also obtained the ground-state energies for $^6$Li for different sets of parameters using both generator A and generator B. These generators are frequency dependent and thus the calculations are not variational in the HO frequency. However, minima in energy can still be found in most cases and we use the minimum of the highest $N_{\rm max}$ as our best approximation. An example is shown in Fig.~\ref{Lifig2} for $\lambda{=}2.0$ fm$^{-1}$ and $N_{gen}{=}10$. Generators A and B give comparable results in this case although the use of the $G_s^B$ flow operator results in somewhat lower energies and a flatter HO frequency dependence.  

Nonetheless, like in most other cases, convergence was not yet achieved and thus we extrapolate to obtain converged energies. We fit the data using an exponential ansatz
\begin{equation}\label{eqn:Fit}
E_{\rm g.s.}=E_0+ae^{-bN_{max}} \; ,
\end{equation}
where $a$, $b$ and $E_0$ are free parameters. $E_0$ is the extrapolated ground-state energy. We note that there exist more sophisticated extrapolation prescriptions~\cite{Coon:2012ab,Furnstahl:2012qg,More2013}, although they are typically applicable in the ultraviolet regime, i.e., in the high HO frequency region. Since we need an extrapolation at a fixed HO frequency that may not guarantee the ultraviolet convergence, we apply the above exponential ansatz applied in various studies in the past~\cite{Navratil:2003ib,Bogner2008}. 

However, the data does not strictly follow an exponential curve and therefore, the fit (\ref{eqn:Fit}) provides a reasonable value when applied only to the data points corresponding to the larger values of $N_{\rm max}$ available. Figure~\ref{Lifig3} shows an example of fitted exponential curves to two data sets where $\hbar\Omega{=}12$~MeV, $\lambda{=}2.0$ fm$^{-1}$ and $N_{\rm gen}{=}10$ or $N_{\rm gen}{=}4$. The sample of extrapolated values, presented in Table~\ref{tbl1} was obtained from a three-points fit at the selected HO frequency, $\hbar\Omega$. The uncertainty is taken from variations of the number of extrapolated points. We also compare to the extrapolated calculation with $G_s{=}T_{\rm rel}$, and, further, to the initial $NN$ potential result based on SRG calculations of Refs.~\cite{Jurgenson:2010wy,Roth:2011ar} with the $3N$-induced interactions included and on the Okubo-Lee-Suzuki calculations of Ref.~\cite{Navratil:2003ib}. 
\begin{table}[t]
\begin{center}
\begin{ruledtabular}
\begin{tabular}{c|cccc}
 Generator  & $\lambda$ [fm$^{-1}$]  &  $N_{gen}$   &  $\hbar\Omega$ [MeV]  &  $E_{\rm g.s.}$ [MeV]  \\
\hline
B & 2.0 & 4 & 12 & -32.82(6) \\
B & 2.0 & 6 & 12 & -32.41(7) \\
B & 2.0 & 8 & 12 & -31.8(1) \\
B & 2.0 & 10 & 12 & -31.2(2) \\
A & 2.0 & 4 & 12 & -32.6(1) \\
A & 2.0 & 10 & 12 & -30.7(3) \\
B & 1.6 & 10 & 12 & -31.6(1) \\
A & 1.6 & 10 & 12 & -31.0(2) \\
B & 1.2 & 4 & 12 & -33.98(2) \\
B & 1.2 & 4 & 14 & -34.06(2) \\
A & 1.2 & 4 & 14 & -34.54(1) \\
B & 1.2 & 10 & 12 & -31.90(8) \\
A & 1.2 & 10 & 12 & -31.2(2) \\
$T_{\rm rel}$ & 2.0 & -& 18 & -32.30(3) \\
Initial $NN$ & - & - & - & -28.0(5) \\
\end{tabular}
\end{ruledtabular}
\caption{Extrapolated ground-state energies of $^6$Li in MeV with uncertainties in parentheses for the three generators and various choices of parameters. The initial $NN$ value is based on results from Refs.~\cite{Jurgenson:2010wy,Roth:2011ar,Navratil:2003ib}. 
}
\label{tbl1}
\end{center}
\end{table}
\begin{table}[t]
\begin{center}
\begin{ruledtabular}
\begin{tabular}{c|cccc}
   & $T_{rel}$ & $G_s^A$ & $G_s^B$ & Expt. \\
\hline
$\mu$ [$\mu_N$] & 0.848(1) & 0.840(3) & 0.841(2) & 0.822 \\
$Q$ [$e$ fm$^2$] & -0.053(18) & -0.054(30) & -0.049(29) & -0.082(2) \\
$B(M1;0_1^+1\rightarrow1_1^+0)$ & 15.10(10) & 14.91(5) & 14.92(5) & 15.43(32) \\
$B(M1;2_1^+1\rightarrow1_1^+0)$ & 0.024(4) & 0.024(2) & 0.025(2) & 0.149(27) \\
$E_{\rm g.s.}$ [MeV] & -32.30(3) & -30.7(3) & -31.2(2) & -31.995 \\
\end{tabular}
\end{ruledtabular}
\caption{Magnetic moment, quadrupole moment, B(M1) transition probabilities (in $\mu_N^2$) and extrapolated ground-state energies of $^6$Li for different flow operators when $\lambda{=}2.0$ fm$^{-1}$, $N_{gen}{=}10$, and $\hbar\Omega{=}12$ MeV. Experimental values are taken from~\cite{Tilley2002}.}
\label{tbl3}
\end{center}
\end{table}
\begin{figure}[tbh]
\begin{center}
\includegraphics[clip=,width=0.45\textwidth]{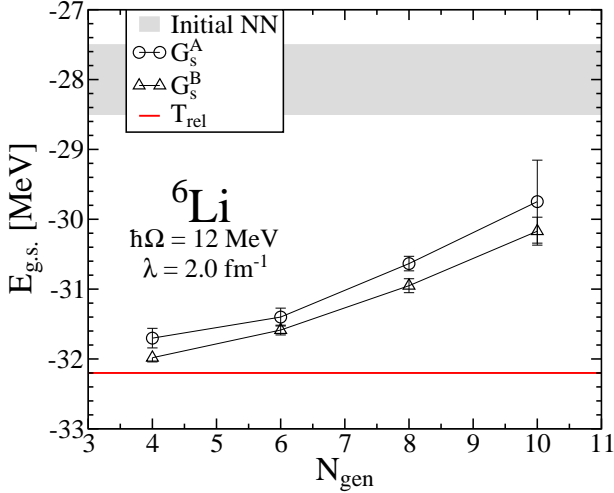}
\end{center}
\caption{(color online) Ground-state energy as a function of $N_{\rm gen}$ for $G_s^A$ and $G_s^B$ with $\hbar\Omega{=}12$ MeV and $\lambda{=}2.0$ fm$^{-1}$. The result obtained with the $G_s{=}T_{\rm rel}$ is shown by the solid line. The shaded band represents the initial $NN$ value with its uncertainty based on results from Refs.~\cite{Jurgenson:2010wy,Roth:2011ar,Navratil:2003ib}.}
\label{Lifig6}
\end{figure}
Overall, the trends are similar to those observed in $^3$H and $^4$He calculations. A larger block model space $P_{\rm gen}$ (higher $N_{\rm gen}$) results in binding energies closer to that of the initial interaction, a lower $N_{\rm gen}$ then takes us closer to calculations with the $G_s{=}T_{\rm rel}$ flow operator. This is also illustrated in Fig.~\ref{Lifig6}.

\begin{figure}[tbh]
\begin{center}
\includegraphics[clip=,width=0.45\textwidth]{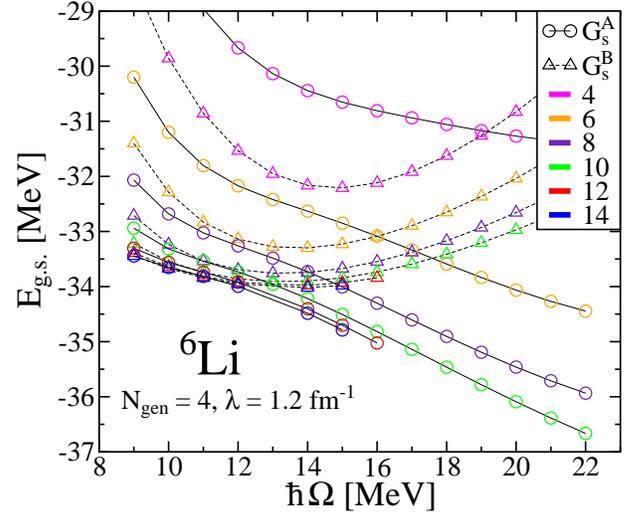}
\end{center}
\caption{(color online) Ground-state energy as a function of HO frequency using generators A and B for different values of $N_{\rm max}$ with $\lambda{=}1.2$ fm$^{-1}$ and $N_{\rm gen}{=}4$.}
\label{Lifig4}
\end{figure}
Calculations discussed so far suggest that the performance of the generators A and B is very similar with the generator A giving somewhat higher binding energies while the convergence is slightly faster with the generator B. However, we observe significant differences in $^6$Li calculations with these two generators at larger $N_{\rm max}$ values when using a small value of $N_{\rm gen}$ ($N_{\rm max}{\gg}N_{\rm gen}$) combined with a smaller lambda. This is illustrated in Fig.~\ref{Lifig4}, where we present $^6$Li ground-state energies  for both generators using $N_{\rm gen}{=}4$ and $\lambda{=}1.2$ fm$^{-1}$ for different basis sizes and a wide range of HO frequencies. While the results obtained with the generator B are similar to those found with higher $\lambda$ and $N_{\rm gen}$ (compare Fig.~\ref{Lifig2}), when generator A is used we observe minima at $N_{\rm max}{=}0$ and $N_{\rm max}{=}2$ that shift fast to the right towards higher HO frequencies with increasing $N_{\rm max}$. For larger values of $N_{\rm max}$ we do not find a minimum in energy in the frequency range up to 22 MeV displayed in the figure. At the same time, we find a dramatic increase of the binding energy with increasing $\hbar\Omega$. In fact the binding energy becomes much larger than that obtained with the standard $G_s{=}T_{\rm rel}$ generator. We believe this to be due to the structural nature of the generator A. Indeed, when using a basis space $N_{\rm max}$ with a dimension larger than the $P_{\rm gen}$ block generator size characterized by the $N_{\rm gen}$, we probe evolved $NN$ matrix elements in the $Q_{\rm gen}$ space. The strong coupling at the boundary of the $P_{\rm gen}$ and $Q_{\rm gen}$ spaces, due to the kinetic operator, induces significant off-diagonal matrix elements beyond the $P_{\rm gen}$ space when the $NN$ potential is evolved to a small $\lambda$ (see Fig.~\ref{fig4}), which most likely contributes to the increase in binding. As demonstrated in Sect.~\ref{Subsec:Potentials}, the generator B does not induce such off-diagonal matrix elements. 

\begin{figure}[tbh]
\begin{center}
\includegraphics[clip=,width=0.45\textwidth]{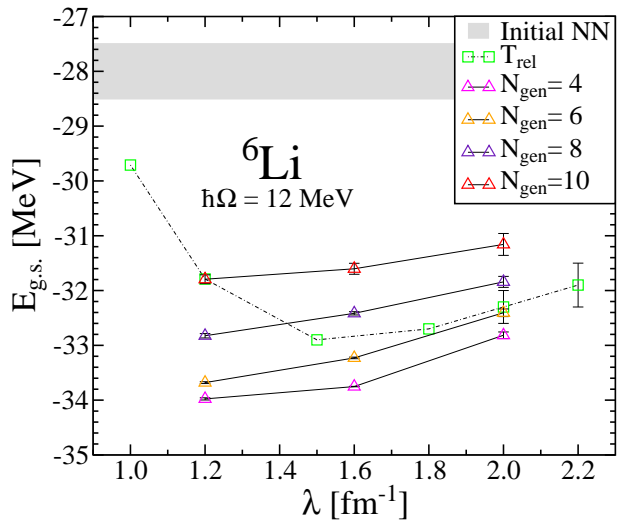}
\end{center}
\caption{(color online) Ground-state energy as a function of $\lambda$ using generators and B for a range of $N_{\rm gen}$ values and $\hbar\Omega{=}12$~MeV. Results obtained with the $G_s{=}T_{\rm rel}$ (connected by the dotted line) are independent of the HO frequency. The shaded band represents the initial $NN$ value with its uncertainty based on results from Refs.~\cite{Jurgenson:2010wy,Roth:2011ar,Navratil:2003ib}.}
\label{Lifig5}
\end{figure}
Finally, we analyzed the lambda dependence of the $^6$Li results. Figure~\ref{Lifig5} shows the extrapolated $^6$Li ground-state energies as a function of $\lambda$ when using generators A and B with different values of $N_{\rm gen}$ and a fixed $\hbar\Omega{=}12$~MeV. It is well known that there is a significant $\lambda$ dependence when using the $G_s{=}T_{\rm rel}$ generator in $NN$-only calculations~\cite{Jurgenson:2010wy,Roth:2011ar}. Using generator A or B does not remove this dependence although it is greatly minimized. This is once the $N_{\rm gen}\hbar\Omega$ is fixed, see the discussion at the end of the Subsection \ref{Subsec:3H+4He}, i.e., we trade the $\lambda$ dependence for the $N_{\rm gen}\hbar\Omega$ dependence, see also Figs.~\ref{Hfig7}, \ref{Hfig6}, \ref{Hfig4} and ~\ref{Lifig6}. We observe that using larger values of $\lambda$ and, in particular, larger values of $N_{\rm gen}$ yields results that are closer to the initial $NN$ interaction value, i.e., the induced many-body interactions are less significant. 

\begin{figure}[tbh]
\begin{center}
\includegraphics[clip=,width=0.45\textwidth]{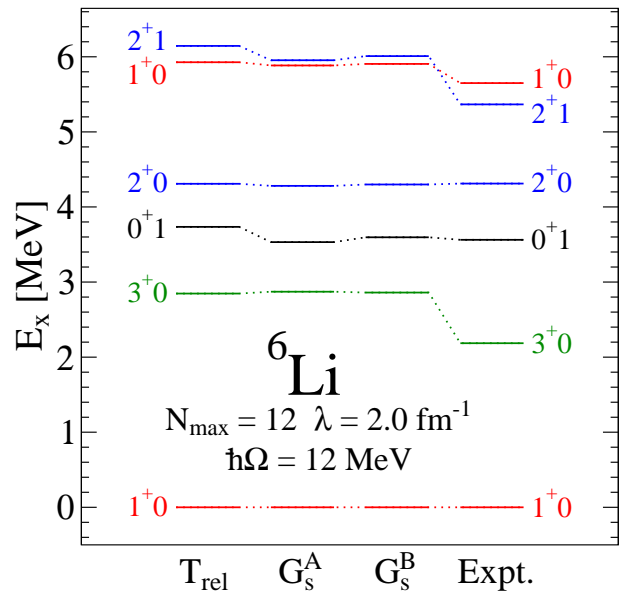}
\end{center}
\caption{(color online) Excited states of $^6$Li in MeV obtained with $G_s{=}T_{rel}$, $G_s^A$, and $G_s^B$. Experimental data from~\cite{Tilley2002} are also shown.}
\label{Lifig7}
\end{figure}
So far we have discussed only ground-state energy results. In Fig.~\ref{Lifig7} we compare $^6$Li excitation energies obtained with the generators A, B and the standard $G_s{=}T_{\rm rel}$ to experiment. We use the values of $N_{\rm gen}{=}10$ and $\lambda{=}2.0$~fm$^{-1}$, best performing in the ground-state calculations (Fig.~\ref{Lifig5}). We find that all the SRG generators give similar excitation spectra that are in a reasonable agreement with experiment. We note that only the ground state of $^6$Li is bound, the excited state are resonances above the $^4$He+$d$ threshold. However, the $3^+ 0$ and the $0^+1$ states are very narrow, therefore we can be confident that our NCSM HO basis calculations are capable of describing these states well. That is not true for the broad $2^+ 0$ and $1^+ 0$ $d$-$^4$He $D$-wave resonances. In Table~\ref{tbl3}, we present the ground-state magnetic and quadrupole moments, and selected B(M1) transition probabilities calculated with the three respective generators. As for the excitation energies, the differences are minimal and the agreement with experiment is quite satisfactory.

\section{Conclusions and outlook}
\label{Sec:Conclusions}

We evolved the chiral $NN$ interaction by SRG transformations using two novel generators with flow operators $G_s$ of block structure in the HO basis. The first generator (A), see Eq.~(\ref{eqn:Ablock}), was motivated by an OLS-type decoupling of the model space $P_{\rm gen}$, defined by a HO cut $N\leq N_{\rm gen}$, and the $Q_{\rm gen}{=}1{-}P_{\rm gen}$ space. The second one (B), see Eq.~(\ref{eqn:Bblock}), was constructed in a way that would leave invariant a Hamiltonian consisting of a kinetic energy plus a $NN$ potential with negligible matrix elements for basis states with $N{>}N_{\rm gen}$. We found that both generators drive the two-body Hamiltonians to block diagonal structure. The evolved $NN$ potentials were then used in NCSM calculations for $^3$H, $^4$He and $^6$Li. We varied the SRG evolution parameter $s$ as well as the cut $N_{\rm gen}$ and compared binding energy results to calculations with the standard generator, $G_s{=}T_{\rm rel}$, as well as to the exact binding energies obtained with the initial chiral $NN$ potential.

We observed that a good convergence comparable to that of the standard $G_s{=}T_{\rm rel}$ generator can be achieved with both block generators if $N_{\rm gen}$ is chosen lower than a reachable $N_{\rm max}$ of the many-nucleon basis. We also observed that unless the $N_{\rm gen}$ is too small (i.e., $\lesssim 6$), the block generators appear to induce weaker many-body forces than the standard $G_s{=}T_{\rm rel}$ generator, i.e., the calculated binding energies are closer to the exact ones when using the block generators. 

When comparing the performance of the two block generators, we saw in $^6$Li calculations that with the first one (A) the HO frequency dependence becomes stronger and at high frequencies the induced many-body forces become significant. This is most likely related to the fact that at large $N_{\rm max}{\gg}N_{\rm gen}$ the $Q$-space part of the $NN$ potential is probed and, further, that the strong coupling at the boundary of the $P_{\rm gen}$ and $Q_{\rm gen}$ spaces due to the kinetic operator, induces some off-diagonal matrix elements beyond the $P_{\rm gen}$ space during the evolution at intermediate values of $\lambda$. Still, the total strength in the off-diagonal blocks decreases monotonically during the evolution.

Overall, our results presented in this study suggest that by using the generator (B) with a selected sufficiently large $N_{\rm gen}$ related to the highest reachable $N_{\rm max}$ by $N_{\rm gen}\sim N_{\rm max}{-}4$, and, further, by selecting a sufficiently small $s$ so that the convergence of the many-nucleon calculation can still be reached, we can reduce the induced many-body force and obtain results closer to the exact ones compared to calculations with the standard $G_s{=}T_{\rm rel}$ generator.

Obviously, the next task is to test the block generators in three-body space also including initial chiral $3N$ interactions and study if the problem of the induced $4N$ interaction will be alleviated. An important issue to explore is the embedding of the evolved two- and three-body interactions in many-body spaces. Also, one may want to test analogous block generators defined in the momentum space rather than in the HO basis as done in this work. A separate important problem to be explored is the evolution of general operators~\cite{Anderson:2010aq,Schuster:2014lga}.

\acknowledgments
Computing support for this work came in part from the LLNL institutional Computing Grand Challenge program and from an INCITE Award on the Titan supercomputer of the Oak Ridge Leadership Computing Facility (OLCF) at ORNL. Support from the Natural Sciences and Engineering Research Council of Canada (NSERC) Grant No. 401945-2011. TRIUMF receives funding via a contribution through the National Research Council Canada.

\bibliographystyle{apsrev}

\begin{thebibliography}{99}

\bibitem{Epelbaum_2009} E. Epelbaum, H.-W. Hammer, and U.-G. Mei�ner, Rev. Mod. Phys. 81, 1773 (2009).

\bibitem{Machleidt:2011zz}
  R.~Machleidt and D.~R.~Entem,
  Phys.\ Rept.\  {\bf 503}, 1 (2011).

\bibitem{Navratil:2000ww}
  P.~Navr\'atil, J.~P.~Vary and B.~R.~Barrett,
  Phys.\ Rev.\ Lett.\  {\bf 84}, 5728 (2000).

\bibitem{Barrett:2013nh}
  B.~R.~Barrett, P.~Navr\'atil and J.~P.~Vary,
  Prog.\ Part.\ Nucl.\ Phys.\  {\bf 69}, 131 (2013).

\bibitem{Hagen:2012fb} 
  G.~Hagen, M.~Hjorth-Jensen, G.~R.~Jansen, R.~Machleidt and T.~Papenbrock,
  Phys.\ Rev.\ Lett.\  {\bf 109}, 032502 (2012).

\bibitem{Hagen:2013nca}
  G.~Hagen, T.~Papenbrock, M.~Hjorth-Jensen and D.~J.~Dean,
  arXiv:1312.7872 [nucl-th].

\bibitem{Binder:2012mk}
  S.~Binder, J.~Langhammer, A.~Calci, P.~Navr\'atil and R.~Roth,
  Phys.\ Rev.\ C {\bf 87}, 021303 (2013).

\bibitem{Binder:2013xaa} 
  S.~Binder, J.~Langhammer, A.~Calci and R.~Roth,
  arXiv:1312.5685 [nucl-th].

\bibitem{Cipollone:2013zma}
  A.~Cipollone, C.~Barbieri and P.~Navr\'atil,
  Phys.\ Rev.\ Lett.\  {\bf 111}, 062501 (2013).

\bibitem{Soma:2013xha} 
  V.~Som\`a, A.~Cipollone, C.~Barbieri, P.~Navr\'atil and T.~Duguet,
Phys. Rev. C {\bf 89}, 061301(R) (2014).

\bibitem{Hergert:2012nb} 
  H.~Hergert, S.~K.~Bogner, S.~Binder, A.~Calci, J.~Langhammer, R.~Roth and A.~Schwenk,
  Phys.\ Rev.\ C {\bf 87}, no. 3, 034307 (2013).

\bibitem{Hergert:2013uja}
  H.~Hergert, S.~Binder, A.~Calci, J.~Langhammer and R.~Roth,
  Phys.\ Rev.\ Lett.\  {\bf 110}, no. 24, 242501 (2013).

\bibitem{Epelbaum:2012qn} 
  E.~Epelbaum, H.~Krebs, T.~A.~L\"ahde, D.~Lee and Ulf-G. Meissner,
  Phys.\ Rev.\ Lett.\  {\bf 109}, 252501 (2012).

\bibitem{Epelbaum:2013paa} 
  E.~Epelbaum, H.~Krebs, T.~A.~L\"ahde, D.~Lee, Ulf-G. Meissner and G.~Rupak,
  Phys.\ Rev.\ Lett.\  {\bf 112}, 102501 (2014).

\bibitem{Lahde:2013uqa} 
  T.~A.~L\"ahde, E.~Epelbaum, H.~Krebs, D.~Lee, Ulf-G.~Meissner and G.~Rupak,
  Phys.\ Lett.\ B {\bf 732}, 110 (2014).

\bibitem{Baroni:2012su}
  S.~Baroni, P.~Navr\'atil and S.~Quaglioni,
  Phys.\ Rev.\ Lett.\  {\bf 110}, 022505 (2013).

\bibitem{Baroni:2013fe}
  S.~Baroni, P.~Navr\'atil and S.~Quaglioni,
  Phys.\ Rev.\ C {\bf 87}, no. 3, 034326 (2013).

\bibitem{Hagen:2012rq}
  G.~Hagen and N.~Michel,
  Phys.\ Rev.\ C {\bf 86}, 021602 (2012).

\bibitem{Ekstrom:2013kea}
  A.~Ekstr\"om, G.~Baardsen, C.~Forss\'en, G.~Hagen, M.~Hjorth-Jensen, G.~R.~Jansen, R.~Machleidt and W.~Nazarewicz {\it et al.},
  Phys.\ Rev.\ Lett.\  {\bf 110}, 192502 (2013).

\bibitem{Glazek93} S. D. Glazek and K. G. Wilson, Phys. Rev. D {\bf 48}, 5863 (1993); {\bf 49}, 4214 (1994).

\bibitem{Wegner94} F. Wegner, Ann. de. Phys. {\bf 506}, 77 (1994); Phys. Rep. {\bf 348}, 77 (2001).

\bibitem{Furnstahl:2012fn}
  R.~J.~Furnstahl,
  Nucl.\ Phys.\ Proc.\ Suppl.\  {\bf 228}, 139 (2012).

\bibitem{Furnstahl:2013oba}
  R.~J.~Furnstahl and K.~Hebeler,
  Rept.\ Prog.\ Phys.\  {\bf 76}, 126301 (2013).

\bibitem{Bogner:2006pc}
  S.~K.~Bogner, R.~J.~Furnstahl and R.~J.~Perry,
  Phys.\ Rev.\ C {\bf 75}, 061001 (2007).

\bibitem{Kehrein:2006ti}
  S.~Kehrein, Springer Tracts Mod. Phys. {\bf 217}, 137 (2006).

\bibitem{Jurgenson:2009qs}
  E.~D.~Jurgenson, P.~Navr\'atil and R.~J.~Furnstahl,
  Phys.\ Rev.\ Lett.\  {\bf 103}, 082501 (2009).

\bibitem{Jurgenson:2010wy}
  E.~D.~Jurgenson, P.~Navr\'atil and R.~J.~Furnstahl,
  Phys.\ Rev.\ C {\bf 83}, 034301 (2011).

\bibitem{Roth:2011ar}
  R.~Roth, J.~Langhammer, A.~Calci, S.~Binder and P.~Navr\'atil,
  Phys.\ Rev.\ Lett.\  {\bf 107}, 072501 (2011).

\bibitem{Roth:2011vt}
  R.~Roth, S.~Binder, K.~Vobig, A.~Calci, J.~Langhammer and P.~Navr\'atil,
  Phys.\ Rev.\ Lett.\  {\bf 109}, 052501 (2012).

\bibitem{Hagen:2013yba} 
  G.~Hagen, T.~Papenbrock, A.~Ekstr\"om, K.~A.~Wendt, G.~Baardsen, S.~Gandolfi, M.~Hjorth-Jensen and C.~J.~Horowitz,
  Phys.\ Rev.\ C {\bf 89}, 014319 (2014).

\bibitem{Entem:2003ft}
  D.~R.~Entem and R.~Machleidt,
  Phys.\ Rev.\ C {\bf 68}, 041001 (2003).

\bibitem{Roth} R. Roth, private communication.

\bibitem{Roth:2010bm}
  R.~Roth, T.~Neff and H.~Feldmeier,
  Prog.\ Part.\ Nucl.\ Phys.\  {\bf 65}, 50 (2010).

\bibitem{Hebeler:2012pr}
  K.~Hebeler,
  Phys.\ Rev.\ C {\bf 85}, 021002 (2012).

\bibitem{Wendt:2013bla}
  K.~A.~Wendt,
  Phys.\ Rev.\ C {\bf 87}, no. 6, 061001 (2013).

\bibitem{Wendt:2011qj}
 K.~A.~Wendt, R.~J.~Furnstahl and R.~J.~Perry,
 Phys.\ Rev.\ C {\bf 83}, 034005 (2011).

\bibitem{Anderson:2008mu}
  E.~Anderson, S.~K.~Bogner, R.~J.~Furnstahl, E.~D.~Jurgenson, R.~J.~Perry and A.~Schwenk,
  Phys.\ Rev.\ C {\bf 77}, 037001 (2008).

\bibitem{Bogner:2002yw}
  S.~Bogner, T.~T.~S.~Kuo, L.~Coraggio, A.~Covello and N.~Itaco,
  Phys.\ Rev.\ C {\bf 65}, 051301 (2002).

\bibitem{Bogner:2003wn}
  S.~K.~Bogner, T.~T.~S.~Kuo and A.~Schwenk,
  Phys.\ Rept.\  {\bf 386}, 1 (2003).

\bibitem{Li:2011sr}
  W.~Li, E.~R.~Anderson and R.~J.~Furnstahl,
  Phys.\ Rev.\ C {\bf 84}, 054002 (2011).

\bibitem{Okubo:1954zz}
  S.~Okubo,
  Prog.\ Theor.\ Phys.\  {\bf 12}, 603 (1954).

\bibitem{Suzuki:1980yp}
  K.~Suzuki and S.~Y.~Lee,
  Prog.\ Theor.\ Phys.\  {\bf 64}, 2091 (1980).

\bibitem{Suzuki:1982aa_JPV}
Suzuki K 1982 {\it Prog. Theor. Phys.} {\bf 68} 246.

\bibitem{Suzuki:1983aa_JPV}
Suzuki K and Okamoto R 1983 {\it Prog. Theor. Phys.} {\bf 70} 439.

\bibitem{Shirokov:2003kk}
  A.~M.~Shirokov, A.~I.~Mazur, S.~A.~Zaytsev, J.~P.~Vary and T.~A.~Weber,
  Phys.\ Rev.\  C {\bf 70}, 044005 (2004)

\bibitem{Navratil:1999pw}
  P.~Navr\'atil, G.~P.~Kamuntavicius and B.~R.~Barrett,
  Phys.\ Rev.\ C {\bf 61}, 044001 (2000).

\bibitem{Anderson:2010aq}
E.~R.~Anderson, S.~K.~Bogner, R.~J.~Furnstahl and R.~J.~Perry,
Phys.\ Rev.\ C {\bf 82}, 054001 (2010).

\bibitem{Schuster:2014lga}
  M.~D. Schuster, S.~Quaglioni, C.~W.~Johnson, E.~D.~Jurgenson and P.~Navr\'atil,
  arXiv:1402.7106 [nucl-th].

\bibitem{Coon:2012ab} 
  S.~A.~Coon, M.~I.~Avetian, M.~K.~G.~Kruse, U.~van Kolck, P.~Maris and J.~P.~Vary,
  Phys. Rev. C {\bf 86}, 054002 (2012).  

\bibitem{Furnstahl:2012qg} 
  R.~J.~Furnstahl, G.~Hagen and T.~Papenbrock,
  Phys.\ Rev.\ C {\bf 86}, 031301 (2012).

\bibitem{More2013} S. N. More, A. Ekstr\"om, R. J. Furnstahl, G. Hagen, and T. Papenbrock, Phys. Rev. C {\bf 87}, 044326 (2013).

\bibitem{Navratil:2003ib} 
  P.~Navr{\'a}til and E.~Caurier,
  Phys.\ Rev.\ C\ {\bf 69}, 014311(2004).

\bibitem{Bogner2008} S. K. Bogner, R. J. Furnstahl, P. Maris, R. J. Perry, A. Schwenk,
and J. P. Vary, Nucl. Phys. A {\bf 801}, 21 (2008).

\bibitem{Tilley2002} D. R. Tilley, C. M. Cheves, J. L. Godwin, G. M. Hale, H. M. Hofmann, J.H. Kelley, C.G. Sheu, H.R. Weller, Nuclear Physics A {\bf 708}, 3 (2002).

\end{thebibliography}

\end{document}